\documentclass[aip, cha,reprint,amsmath,amsfonts]{revtex4-1}

\usepackage{amssymb}
\usepackage{amsmath}
\usepackage{bm}
\usepackage{paralist}
\usepackage{graphicx}
\usepackage{enumerate}
\usepackage{tikz}
\usepackage{graphicx}
\usepackage{verbatim}
\usepackage{etoolbox}

\usetikzlibrary{matrix,positioning,fit}

\usepackage{hyperref}
\hypersetup{
	colorlinks,
	citecolor=blue,
	filecolor=blue,
	linkcolor=blue,
	urlcolor=blue,
	pdfproducer={}
}

\let\bbordermatrix\bordermatrix
\patchcmd{\bbordermatrix}{8.75}{4.75}{}{}
\patchcmd{\bbordermatrix}{\left(}{\left[}{}{}
\patchcmd{\bbordermatrix}{\right)}{\right]}{}{}

\newlength{\Mylen}

\usepackage{ragged2e}
\usepackage{caption}
\usepackage{ragged2e}
\DeclareCaptionJustification{justified}{\justifying}
\usepackage{soul}	%used for striking through text - \st{text} 

\usepackage [autostyle, english = american]{csquotes}
\MakeOuterQuote{"}

\usepackage[normalem]{ulem}

\usepackage[percent]{overpic}

\usepackage{xcolor}

\usepackage{cprotect}

\begin{document} 
\title{Fractal generation in a two-dimensional active-nematic fluid}

\author{Kevin A.~Mitchell}
\email{kmitchell@ucmerced.edu}

\author{Amanda J.~Tan}

\author{Jorge Arteaga}

\author{Linda S.~Hirst}

\affiliation{Physics Department, University of California, Merced, CA
  95344, USA}

\date{\today}
	
\begin{abstract}
  Active fluids, composed of individual self-propelled agents, can
  generate complex large-scale coherent flows.  A particularly
  important laboratory realization of such an active fluid is a system
  composed of microtubules, aligned in a quasi-two-dimensional (2D)
  nematic phase, and driven by ATP-fueled kinesin motor proteins.
  This system exhibits robust chaotic advection and gives rise to a
  pronounced fractal structure in the nematic contours.  We
  characterize such experimentally derived fractals using the power
  spectrum and discover that the power spectrum decays as $k^{-\beta}$
  for large wavenumbers $k$.  The parameter $\beta$ is measured for
  several experimental realizations.  Though $\beta$ is effectively
  constant in time, it does vary with experimental parameters,
  indicating differences in the scale-free behavior of the
  microtubule-based active nematic.  Though the fractal patterns
  generated in this active system are reminiscent of passively
  advected dye in 2D chaotic flows, the underlying mechanism for
  fractal generation is more subtle.  We provide a simple, physically
  inspired mathematical model of fractal generation in this system
  that relies on the material being locally compressible, though the
  total area of the material is conserved globally.  The model also
  requires that large-scale density variations be injected into the
  material periodically.  The model reproduces the power spectrum
  decay $k^{-\beta}$ seen in experiments.  Linearizing the model of
  fractal generation about the equilibrium density, we derive an
  analytic relationship between $\beta$ and a single dimensionless
  quantity $r$, which characterizes the compressibility.
\end{abstract}

\maketitle

\begin{quotation}
Active fluids are out-of-equilibrium systems that
  exhibit spontaneous flow dynamics as they consume energy locally to
  generate material stresses~\cite{Marchetti13}. There are many
  fascinating examples of active fluids in nature that cross a wide
  range of length scales, from bacterial
  colonies~\cite{DellArciprete18,You18,Li19} and cellular
  sheets~\cite{Saw17,Kawaguchi17} to bird flocks~\cite{Toner95} and
  insect swarms~\cite{Buhl06}.  We study a laboratory model of an
  active fluid with nematic ordering~\cite{Sanchez12}, which is
  manifested by a striped pattern whose orientation varies throughout
  the fluid.  This system exhibits spontaneous chaotic advection on
  submilimeter scales, thus making it an interesting paradigm for
  studies of chaotic dynamics.  We show that the density fluctuations
  inherent in the striped patterning exhibit a fractal scaling.  We
  then provide a simplified mathematical model for the generation of
  this fractal structure.  This model shows how the material
  properties can be quantitatively linked to the decay of the power
  spectrum on small scales.  We believe this provides an interesting
  new model of fractal generation in addition to providing insight
  into the physics of active nematics.
\end{quotation}

\section{Introduction}

We examine a quasi-two-dimensional (2D) synthetic active nematic
fluid, composed of biologically derived subunits~\cite{Ndlec97}, in
which the rod-like microtubule subunits are driven relative to each
other by the action of kinesin motor proteins.  In recent years, this
system has become an important prototype for active fluids, with
numerous experimental and theoretical
studies~\cite{Sanchez12,Henkin14,Giomi15,DeCamp15,Guillamat16,Doostmohammadi17,Guillamat17,Shendruk17,Lemma19,Tan19}.
A key characteristic of this fluid is the emergence of mobile
topological defects; points where the orientational order of the
nematic phase breaks down. Positive and negative defects occur
spontaneously as a result of the active flows, separate from each
other and ultimately annihilate with other defects of their opposite
charge in the fluid.  In recent work, our group analyzed the motion of
these defects in the context of chaotic mixing~\cite{Tan19}, revealing
that the defects can be treated as virtual stirring rods, which braid
around each other, driving advective flows that stretch and fold the
material.

\begin{figure}
\includegraphics[width = 1\columnwidth]{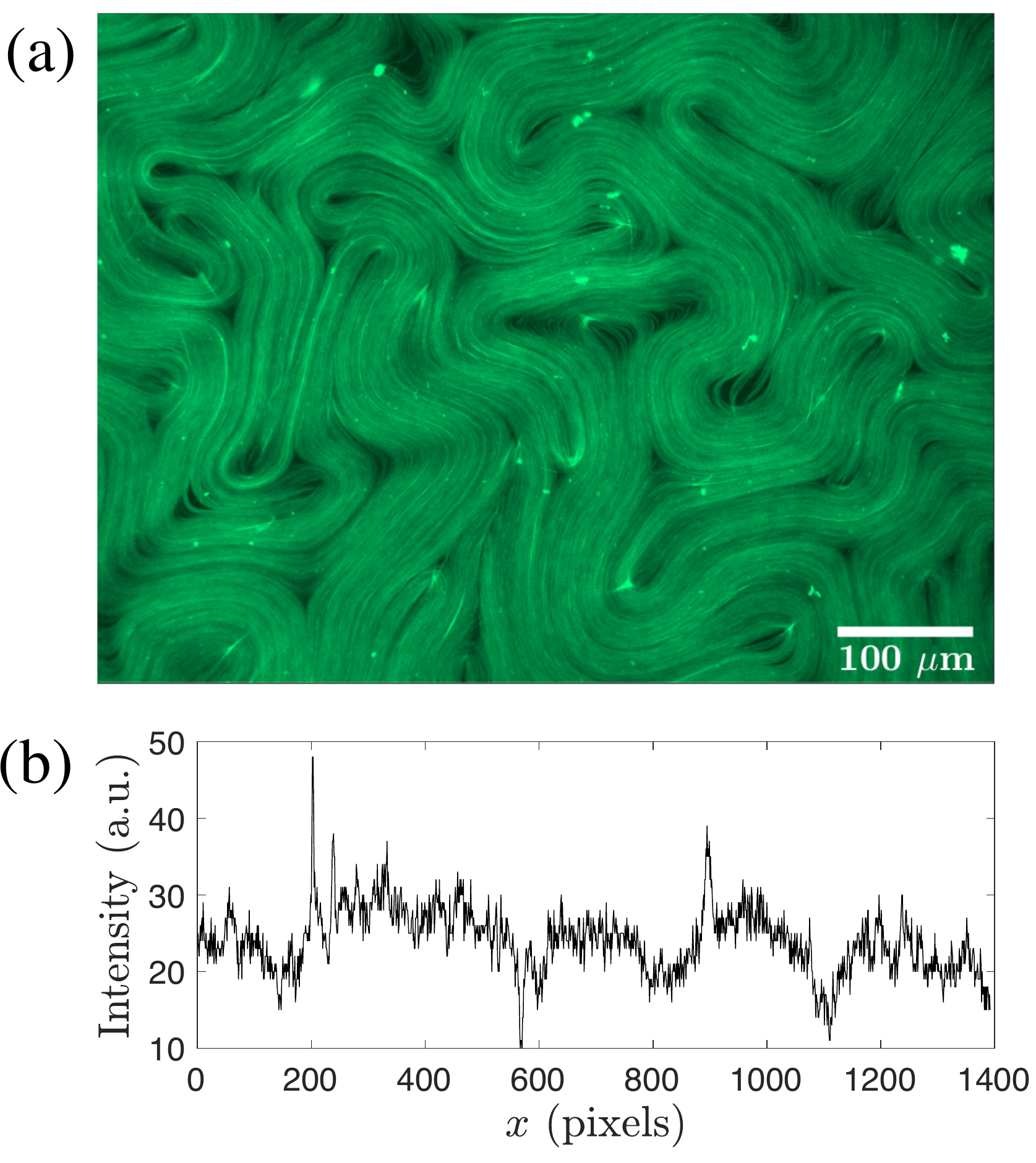} 
\caption{\label{fig8} a) A fluorescence microscopy image of the microtubule
  system.  b) The intensity across the top row of panel a.}
\end{figure}

One of the most striking features of 2D microtubule-based active
nematics is the visually distinctive patterns of folded dark and light
bands of varying intensity (Fig.~\ref{fig8}a).  These folded patterns
appear fractal, with finer structure apparent at smaller length
scales.  These fractal patterns are the subject of the current paper.
The fractal analysis of images has a rich history across numerous
disciplines, with many quantitative measures of fractal behavior
introduced and
utilized~\cite{Peitgen88,Soille96,Lopes09,Bies16,Nayak19}.  Fractal
dimension is a well known tool and various definitions are commonly
used.  Here we choose to investigate the closely related decay of the
power spectrum at large wavenumbers, i.e. small length scales.  We
analyze experimental images of active nematics, using data sets first
reported in Ref.~\onlinecite{Tan19}.  We find that power, i.e. the
norm-squared of the Fourier coefficients, scales as a power-law
$k^{-\beta}$ in the wavenumber $k$.  This power-law behavior begins at
the active length scale of the system, roughly the spacing between
topological defects, and continues down to smaller scales, eventually
being dominated by pixel noise.  In a log-log plot, power versus $k$
displays a strikingly linear behavior over a wide range of $k$ values,
from which we extract $\beta$.  The value of $\beta$ is roughly
constant in time.

Fractal patterns are well known to occur within the passive advection
of material in a traditional 2D
fluid~\cite{Tel05,Aguirre09,Lai10,Aref89,Muzzio92b,Pentek95}.  This is
evident on large planetary scales, e.g. plankton blooms in the
ocean~\cite{Tel05,Sandulescu07}, or on small laboratory scales,
e.g. microfluidic devices~\cite{Truesdell03,Phelan08}.  However, in a
confined, incompressible flow, such fractal behavior is transient;
eventually the system becomes well mixed and the fractal pattern
washes out~\cite{Muzzio92b}.  Persistent fractal behavior can arise in
an open incompressible flow, when a constant stream of impurity enters
and then exits a chaotic mixing region.  However, the fractal
structure witnessed in the active microtubule-based nematic is both
persistent and confined.  Furthermore, the microtubule system is also
incompressible, or area-preserving, at least when averaged over a
large enough domain.  On smaller length scales, laboratory videos do
show local compression of microtubule bundles and the creation and
expansion of small voids.  (See the online supplemental video.)
Despite this, most continuum-based simulations assume local
incompressibility from the start~\cite{Giomi15,
  Doostmohammadi17,Shendruk17}.

We present a simple model for the creation of fractal structure that
does not assume incompressibility on small length scales.  Area is,
however, preserved at the active length scale.  Similarly, density is
not constant; the patterns seen in the active nematic are, after all,
due to local variations in density.  Our model contains three distinct
steps: 1) extension of the material in the direction of the director
field, i.e. along the microtubule bundles, and compression in the
perpendicular direction; 2) folding of the material back over itself
in a horseshoe pattern; and 3) creation of new large-scale density
fluctuations at the active length scale.  A critical aspect of this
model is that the compressibility in Step 1 varies with density.  We
find that this simple model generates $k^{-\beta}$ scaling in the
power spectrum with a well defined $\beta$ value.

To better understand the model's consequences, we conduct an analysis
of the fractal formation to first order in the density fluctuations.
Within this linear analysis, we analytically compute $\beta$ and find
that it depends solely on a dimensionless parameter $r$, that
characterizes the compressibility.  The fractal behavior is lost if
the system is incompressible.

This model is not intended to be a quantitatively faithful simulation
of all the physics of a microtubule-based active nematic.  Rather it
is a reduced model designed to highlight the key physical processes
necessary to create fractal structure.  The lessons learned here could
be incorporated into more quantitatively accurate numerical
simulations in the future.

Note that our work should be distinguished from prior studies of
scale-free behavior in active nematics~\cite{Giomi15,Alert20}.  These
studies assume uniform density, so are not sensitive to the density
fluctuations considered here.  Instead, they consider the decay in
the energy and enstrophy spectra, which have a universal power-law
behavior at small scales.

This paper is organized as follows.  Section~\ref{sec:experiments}
presents our results on the spectral decay of experimental images of
the active nematic material.  Section~\ref{sec:model} explains our
three-step model of fractal generation.  Section~\ref{sec:linear}
linearizes the model in the density fluctuations.
Section~\ref{sec:incompressible} discusses the special case in which
Step 1 of our model is incompressible but Step 3 is not.  Conclusions
are in Sect.~\ref{sec:conclusions}.

\section{Experimentally measured power-spectrum decay of microtubule-based
  active nematics}

\label{sec:experiments}

The analysis here is based on the experimental data presented in
Ref.~\onlinecite{Tan19}, where a complete description of the experimental
technique is given.  In brief, a quasi-2D layer of microtubules is
suspended at an oil-water boundary.  The microtubules are at high
density and form bundles due to the use of a depletion agent
(polyethylene glycol).  Kinesin molecular motors, joined together in
clusters, serve to cross-link the microtubules.  ATP is added to the
solution to power the molecular motors, which walk along the
microtubules from their negative end to their positive end.
Neighboring microtubules of opposite polarity are pushed in opposing
directions by the motor clusters to produce the active stress in the
material.  This local stress produces large-scale dynamics in the 2D
microtubule layer.  The system is imaged using fluorescence
microscopy.  See Fig.~\ref{fig8}a and the online supplemental video.
Experiments were performed with six different ATP concentrations.  The
higher the ATP concentration, the more activity is induced in the
system and the faster the system evolves, assuming all other factors
are held constant.

\begin{figure}
\includegraphics[width = 1\columnwidth]{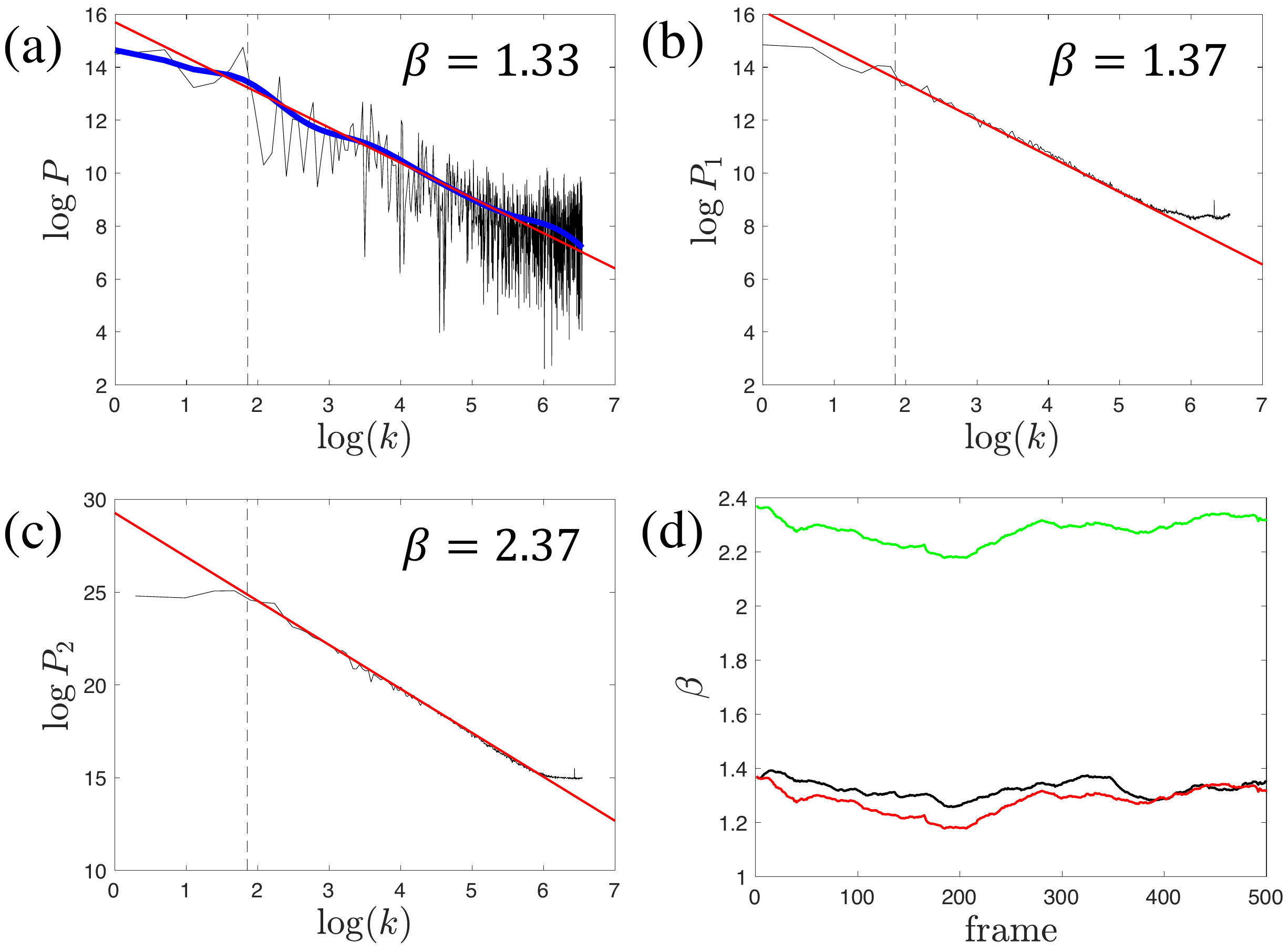} 

\caption{\label{fig7} a) Power spectrum (black) of Fig.~\ref{fig8}b.
  The blue curve is the power spectrum averaged in $\log k$ space
  using a Gaussian of width $\sigma = 0.4$.  (See Appendix.)  The red
  line is a linear fit to this average.  The vertical dashed line
  denotes the $k$ value corresponding to the velocity autocorrelation
  length.  Note that the scaling is chosen so that $k = L_0/\lambda$
  for wavelength $\lambda$, where $L_0$ is the width of the image; b)
  The 1D power spectrum averaged over all rows of the image in
  Fig.~\ref{fig8}a; c) The angle-averaged 2D power spectrum of the
  image in Fig.~\ref{fig8}a; d) The 1D (black) and 2D (green) $\beta$
  values as a function of time.  The red curve is the green curve
  shifted down by 1.  All data in these figures were taken at an ATP
  concentration of $50 \mu$M.}
\end{figure}

We analyse the power spectrum of the resulting experimental images
$f(x,y)$ in two ways.  In the first technique, we compute the 1D
Fourier transform $\tilde{f}(k_x;y)$ of each row; here $y$ labels the
row and $k_x$ is the $x$ wave vector.  The power spectrum of row $y$
is then the norm-squared of the Fourier coefficients versus
$k = |k_x|$, i.e.  $P(k;y) = |\tilde{f}(k;y)|^2$.  For example,
Fig.~\ref{fig8}b shows the intensity of the top row in
Fig.~\ref{fig8}a; Fig.~\ref{fig7}a shows a log-log plot of the
resulting power spectrum (black).  To better see the spectral decay,
we smooth the power spectrum in $\log k$ as discussed in the Appendix.
This average is shown as the smooth blue curve.  Finally, the blue
curve is fit to a linear decay, shown in red, with slope $-1.33$.  The
vertical dashed line marks the value $k = L_0/\ell$, where $L_0$ is
the width of the image and $\ell$ is the active length scale, defined
here as the velocity autocorrelation length~\cite{Lemma19,Hemingway16}
of the active nematic, computed in Ref.~\onlinecite{Tan19}.  Note that the
linear fall off begins at $k$ values larger than approximately
$L_0/\ell$.

The initial (black) power spectrum in Fig.~\ref{fig7}a has large
fluctuations, which we can reduce by averaging $P(k;y)$ over all the
rows $y$ of the image, producing the averaged 1D power spectrum
$P_1(k)$.  This is shown as the black power spectrum in
Fig.~\ref{fig7}b, where the fluctuations have diminished dramatically.
The red fit line is obtained as in Fig.~\ref{fig7}a, i.e. the power
spectrum is first smoothed in $\log k$ (not shown in
Fig.~\ref{fig7}b), and then the red line is fit to this average.  The
resulting slope is $-1.37$, comparable to that of Fig.~\ref{fig7}a.

The second technique for computing the spectral decay is to first take
the 2D Fourier transform $\tilde{f}(k_x,k_y)$ of the entire image (in
practice, a square subset of the image).  Then
$P(k_x,k_y) = |\tilde{f}(k_x,k_y)|^2$ is averaged over angle to
produce the 2D power spectrum
\begin{equation}
P_2(k) = \frac{1}{2 \pi} \int_0^{2 \pi} P(k \cos(\phi), k \sin(\phi))
\,  d\phi,
\label{r21}
\end{equation}
which is shown in Fig.~\ref{fig7}c.  Its fluctuations are 
roughly comparable to that of averaging over the rows
(Fig.~\ref{fig7}b).  The data is again smoothed in $\log k$ and
subsequently fit to the red line, with slope $-2.37$.

Note that these power spectra show that there is a single active
length scale, which can be defined as the velocity autocorrelation
length $\ell$, as done here.  This is consistent with what is known
from prior publications, e.g. Ref.~\onlinecite{Lemma19}.  On scales smaller
than the active length scale, the density distribution appears scale
free.

\begin{figure}
\includegraphics[width = .7\columnwidth]{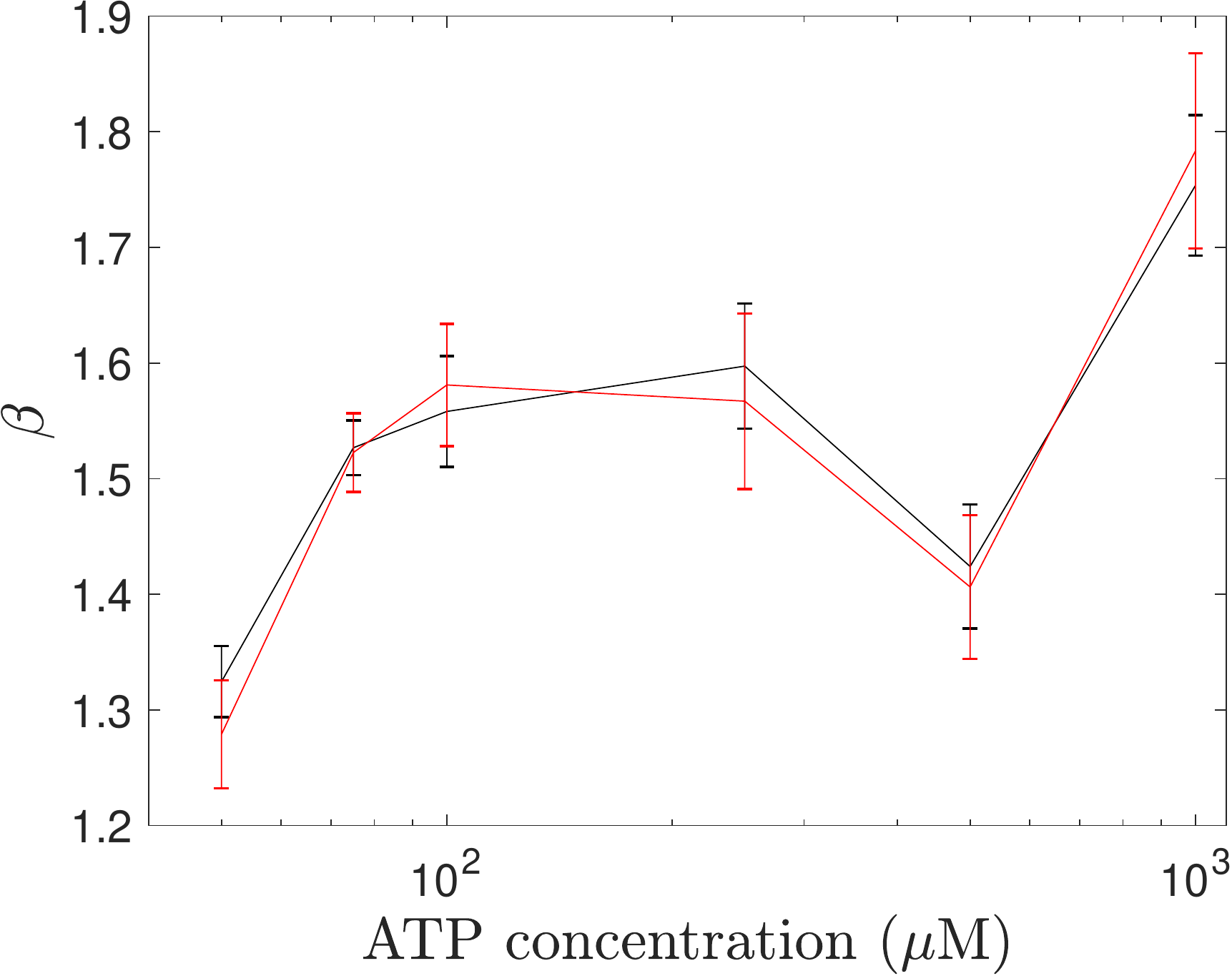} 
\caption{\label{fig9} The $\beta$ value averaged over all 500 frames
  of the active nematic video for experiments performed at six
  different ATP concentrations: 50, 75, 100, 250, 500, 1000 $\mu$M.
  The black data is computed from $P_1$ and the red data from $P_2$,
  shifted down by 1.  Error bars are the standard deviations of
  $\beta$ over all frames.}
\end{figure}

 We now investigate the time-dependence of the $\beta$'s.
Figure~\ref{fig7}d shows $\beta$ computed from $P_1$ (black) and $P_2$
(green) for each of 500 experimental frames.  There is modest
variation in each graph.  Note that the two curves differ by about 1.
This is made clearer by the red curve in Fig.~\ref{fig7}d, which is
the green curve shifted down by 1.  The reason for this correspondence
can be understood as follows.  Suppose the original image $f(x,y)$
were independent of $y$, consisting only of vertical stripes.  Then
$\tilde{f}(k_x, k_y)$ could be written as $\tilde{f}(k_x; y) \delta_{k_y 0}$, i.e. the 2D
Fourier transform depends on $k_x$ just as the 1D Fourier transform of
any (equivalent) row of the image and is nonzero only when $k_y = 0$,
since the image is constant in $y$.  If $P(k_x;y)$ falls off like
$k^{-\beta}$, then the angular average of $P(k_x, k_y)$ is one factor
of $k$ smaller due to dividing by the circumference $2 \pi k$.  The
close agreement between the red and black curves confirms the validity
of our analysis.

 Separate experiments were run for six different ATP concentrations.
Figure~\ref{fig9} shows the $\beta$ value averaged over all 500 frames
for each ATP concentration.  The black data is computed from $P_1$ and
the red data from $P_2$, shifted down by one.  Error bars are the
standard deviation of $\beta$ over the 500 frames.  Note that the
$P_1$ and $P_2$ data are consistent across all six experimental runs.
However, there are significant variations in $\beta$ from one
experiment to the next, indicating that the nature of the scale-free
behavior is not universal.

\section{Model of fractal generation}

\label{sec:model}

\subsection{Step 1: One-dimensional model of compression dynamics}

\begin{figure}
\includegraphics[width = 1\columnwidth]{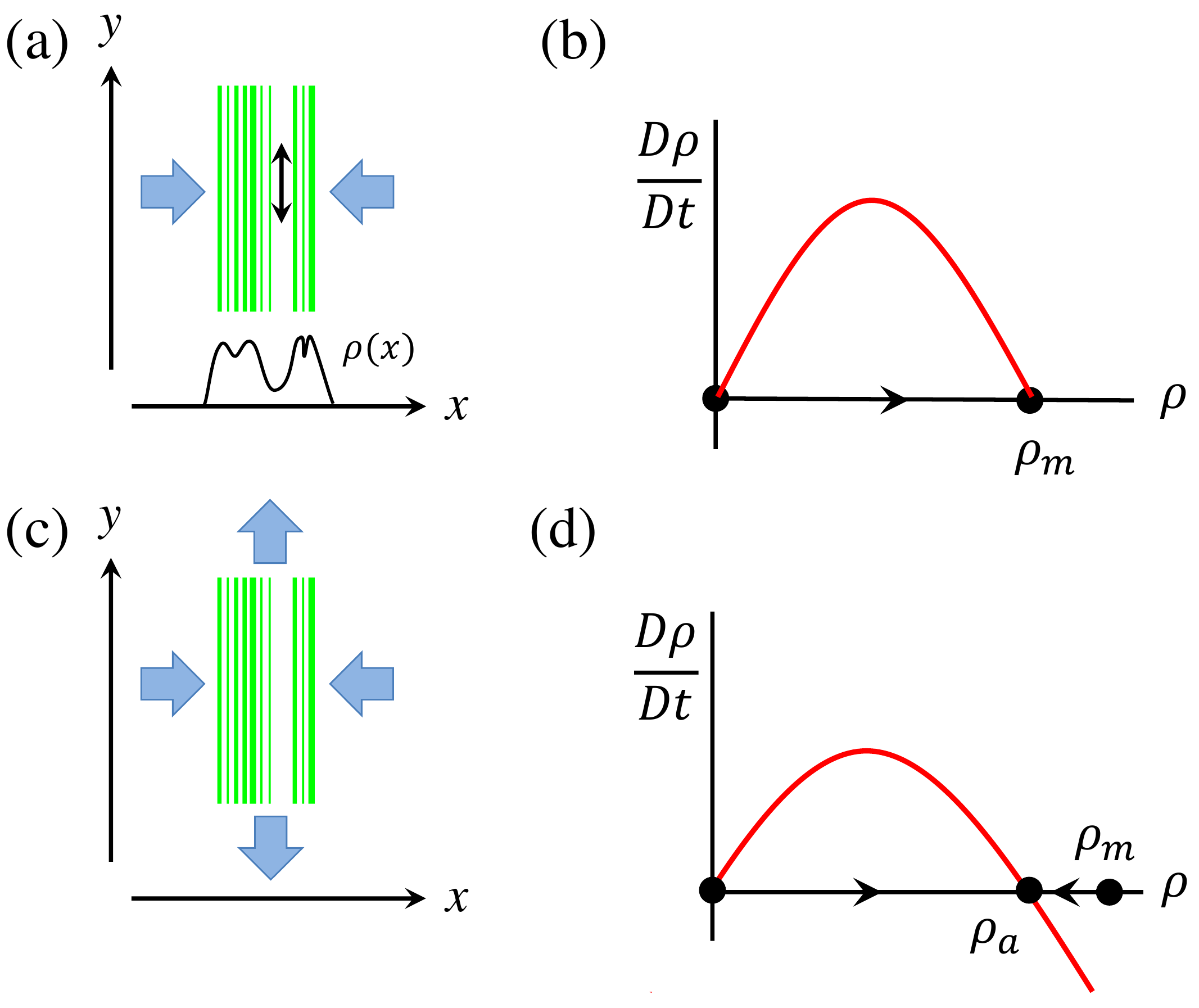} 
\caption{\label{fig10} a) The microtubules (green) and directors
  (black double arrow) are aligned in the $y$ direction.  The
  microtubule density $\rho$ depends only on $x$.  The microtubules
  are compressed in the $x$ direction.  b) The time-rate-of-change of
  the density in the Lagrangian frame.  The density is driven toward
  the stable fixed point $\rho_m$.  c) Stretching in the $y$ direction
is added to the model.  d)  The density is driven toward the new
stable fixed point $\rho_a < \rho_m$.}
\end{figure}

We consider a model of the material in which the
microtubule bundles, and hence the directors, point along the
$y$ direction everywhere, and in which the density $\rho$ depends
solely on $x$.  See Fig.~\ref{fig10}a.  We assume that this material
is under a constant stress perpendicular to the directors and that the
response of the material depends solely on the density $\rho(x)$.
Specifically, the relative change in length $d L/L$ of any interval
$[x_0, x_1]$, where $L = x_1-x_0$, over time $dt$ depends solely on
the average density $\bar{\rho}$ of the interval, i.e.
\begin{align}
\frac{d L}{dt} & = g(\bar{\rho}) L, \label{r1} \\
\bar{\rho} &= \frac{1}{L} \int_{x_0}^{x_1} \rho(x) dx,
\end{align}
where $g$ is the compression rate for a given average density.  The
assumption that Eq.~(\ref{r1}) applies to \emph{all} intervals is
equivalent to $g$ being an affine function, which we write as  
% (*  explain? *)
%
\begin{equation}
g(\bar{\rho}) = -\alpha(1 - \bar{\rho}/\rho_m),
\label{r28}
\end{equation}
where $\alpha$ and $\rho_m$ are constants.  We take $\rho_m > 0$ and
$\alpha > 0$, so that the material becomes less compressible at higher
density, becoming incompressible for the given stress when the average
density is $\rho_m$.  Using $v(x,t)$ for the velocity of the material
at position $x$ and time $t$, Eqs.~(\ref{r1})--(\ref{r28}) imply
\begin{align}
\frac{1}{L} \int_{x_0}^{x_1} \frac{\partial v}{\partial x} dx 
& = \frac{1}{L}[v(x_1) - v(x_0)] = \frac{1}{L} \frac{d L}{dt} =
  g(\bar{\rho})  \nonumber \\
& = -\alpha + \frac{\alpha}{\rho_m}
\frac{1}{L} \int_{x_0}^{x_1} \rho(x) dx \nonumber \\
& =  \frac{1}{L}  \int_{x_0}^{x_1}   -\alpha \left(1 -  \frac{\rho(x)}{\rho_m}\right)  dx, 
\end{align}
from which we find
\begin{equation}
\frac{\partial v}{\partial x} = -\alpha \left( 1 - \frac{\rho(x)}{\rho_m} \right).
\label{r2}
\end{equation}
Applying the continuity equation 
\begin{equation}
\frac{\partial}{\partial t} \rho + 
\frac{\partial}{\partial x} (v \rho) = 0,
\end{equation}
we find
\begin{equation}
\frac{D}{Dt} \rho(x,t) = \frac{\partial}{\partial t} \rho + v
\frac{\partial}{\partial x} \rho = -\rho \frac{\partial}{\partial x} v
=  \alpha \rho \left( 1 - \frac{\rho}{\rho_m}\right).
\label{r3}
\end{equation}
Here  $D/Dt$ is the advective derivative.  Note that the quadratic
nature of Eq.~(\ref{r3}) produces two fixed points for $\rho$ in the
Lagrangian frame, one at $\rho = 0$
and one at $\rho = \rho_m$; these  are linearly unstable and stable,
respectively.  See Fig.~\ref{fig10}b.   Thus, a typical parcel of material will be driven
toward the maximum density $\rho_m$ as it is compressed; a block of
material already at density $\rho_m$ is incompressible.  Note that the
combination of Eqs.~(\ref{r2}) and (\ref{r3}) yields an
integro-differential equation for the density $\rho(x,t)$, which can
be solved numerically.  % (* explain more? *)

We next incorporate into the dynamics stretching in the $y$-direction,
given by the rate $\gamma>0$, which is taken to be constant in
position and time and independent of density.  See Fig.~\ref{fig10}c.
This does not change Eq.~(\ref{r2}), but modifies Eq.~(\ref{r3}) to be
\begin{align}
\frac{D}{Dt} \rho(x,t) &= \frac{\partial}{\partial t} \rho + v
\frac{\partial}{\partial x} \rho 
=  \rho ( \alpha - \gamma - \alpha \frac{\rho}{\rho_m}) \nonumber \\
& =  \alpha \rho \left( \frac{ \rho_a - \rho}{\rho_m} \right),
\label{r4}
\end{align}
where 
\begin{equation}
\rho_a = \rho_m (\alpha - \gamma)/\alpha < \rho_m.
\label{r27}
\end{equation}
See Fig.~\ref{fig10}d.  With this modification, a parcel of material
is driven toward the steady state density $\rho_a$ instead of
$\rho_m$.  

Note that the limit $\rho_m = \infty$, with $\alpha$ and $\gamma$
constant, removes the density dependence from Eqs.~(\ref{r2}) and
(\ref{r4}).  Density then increases at the uniform rate of
$\alpha - \gamma$ everywhere.  This behavior is identical to the
compression-stretching step for both the baker and Lozi maps, and it
is well known that both of these maps have a fractal attractor, with a
singular density distribution, but only when the system is
contracting, i.e. $\alpha - \gamma > 0$.  This is not an accurate
model for the microtubule-based system, which must be area-preserving
on average.  One could instead consider the incompressible limit of
Eq.~(\ref{r4}) by setting $\alpha = \gamma$.  Though completely
area-preserving dynamics will not produce fractal structure, it is
possible to produce fractal structure if Step 1 is area preserving and
it is combined with density fluctuations introduced in Step 3.  (See
Sect.~\ref{sec:densityvariations} below.)  This special case is
considered in Sect.~\ref{sec:incompressible}.  Until then, we take
$\rho_m$ to be finite.

Though area is not conserved locally in the microtubule system, we do
require that area be conserved across the entirety of the material, as
noted above.  This means that the stretching rate must be balanced by
the compression rate, i.e.  $\gamma = -g(\bar{\rho}_T)$, where
$\bar{\rho}_T$ is the average density of the entire block of material.
Since $\gamma$ is assumed constant in time, $\bar{\rho}_T$ must also
be constant in time.  Equation~(\ref{r4}) then implies that
\begin{equation}
\bar{\rho}_T = \rho_a.
\label{r29}
\end{equation}
The ``a'' subscript stands for ``average''.

By an appropriate choice of length and time scales, we can set
$\alpha = \rho_m = 1$ so that Eqs.~(\ref{r2}) and (\ref{r4}) become
\begin{align}
  \frac{D}{Dt} \rho(x,t) & = \frac{\partial}{\partial t} \rho + v
  \frac{\partial}{\partial x} \rho 
  =  \rho ( r - \rho), \label{r5} \\
\frac{\partial v}{\partial x} & = \rho - 1,  \label{r6}
\end{align}
where 
\begin{equation}
r = \rho_a/\rho_m \le 1
\label{r26}
\end{equation}
is dimensionless.  Equation~(\ref{r29}) for the  density averaged
over the entire material becomes
\begin{equation}
\bar{\rho}_T = r.
\label{r22}
\end{equation}
Note that there is no incompressible limit of Eqs.~(\ref{r5}) and
(\ref{r6}), since the rescaling assumes $\rho_m$ is finite. 

\subsection{Step 2: Folding dynamics}

After the system has been compressed to one-half its original width,
we introduce a fold.  In the one-dimensional model, this means that
the system returns  to its original width $L_0$, with density 
\begin{equation}
\rho(x,t) = 
\left\{
\begin{array}{ll}
\rho(x,t) & 0 < x < L_0/2, \\
\rho(L_0/2-x,t) & L_0/2 < x < L_0.
\end{array}
\right.
\label{r10}
\end{equation} 
An alternative perspective here is to extend $\rho(x,t)$ to an
$x$-periodic function with wavelength $L_0$ that is also an even
function in $x$.  In this case, the function $\rho(x,t)$ will
naturally evolve into the form of Eq.~(\ref{r10}) once the original
interval is compressed to half its length.  Thus the fold step is
naturally incorporated into the compression step.

To implement this step, we compute the time $T$ to compress the entire
block of material from length $L_0$ to an arbitrary length $L$.
First, assuming without loss of generality that the position of the
left edge of the block is fixed at $x=0$, the velocity at the right
edge $x = L$ is computed as
\begin{align}
\frac{dL}{dt} & = \int_0^L \frac{\partial v}{\partial x} dx 
= - \int_0^L  ( 1 - \rho) dx  =  -( 1 - r) L,
\end{align}
using Eqs.~(\ref{r6}) and (\ref{r22}).  Hence, the time to compress
the block from $L_0$ to $L$ is
\begin{equation}
T = \frac{ \log(L_0/L)}{1 - r} =  \frac{ \log(2)}{ 1 - r },
\label{r11}
\end{equation}
where in the final step we have set $L = L_0/2$.

\subsection{Step 3: Imposing large-scale density variations}

\label{sec:densityvariations}

Immediately after the folding step we impose a multiplicate density
variation $R(x)$ and normalize the density so that the  average
density remains $r$.  Explicitly,
\begin{equation}
\rho(x) \mapsto \frac{R(x) \rho(x)}{\langle R(x) \rho(x) \rangle} r,
\label{r12}
\end{equation}
where the angled brackets denote the average over $x$.  Consistent
with the viewpoint that $\rho(x)$ is an even, $L_0$-periodic function,
we similarly extend $R(x)$ to an even, $L_0$-periodic function. 

\subsection{Fractal generation}

\begin{figure}
\includegraphics[width = 1\columnwidth]{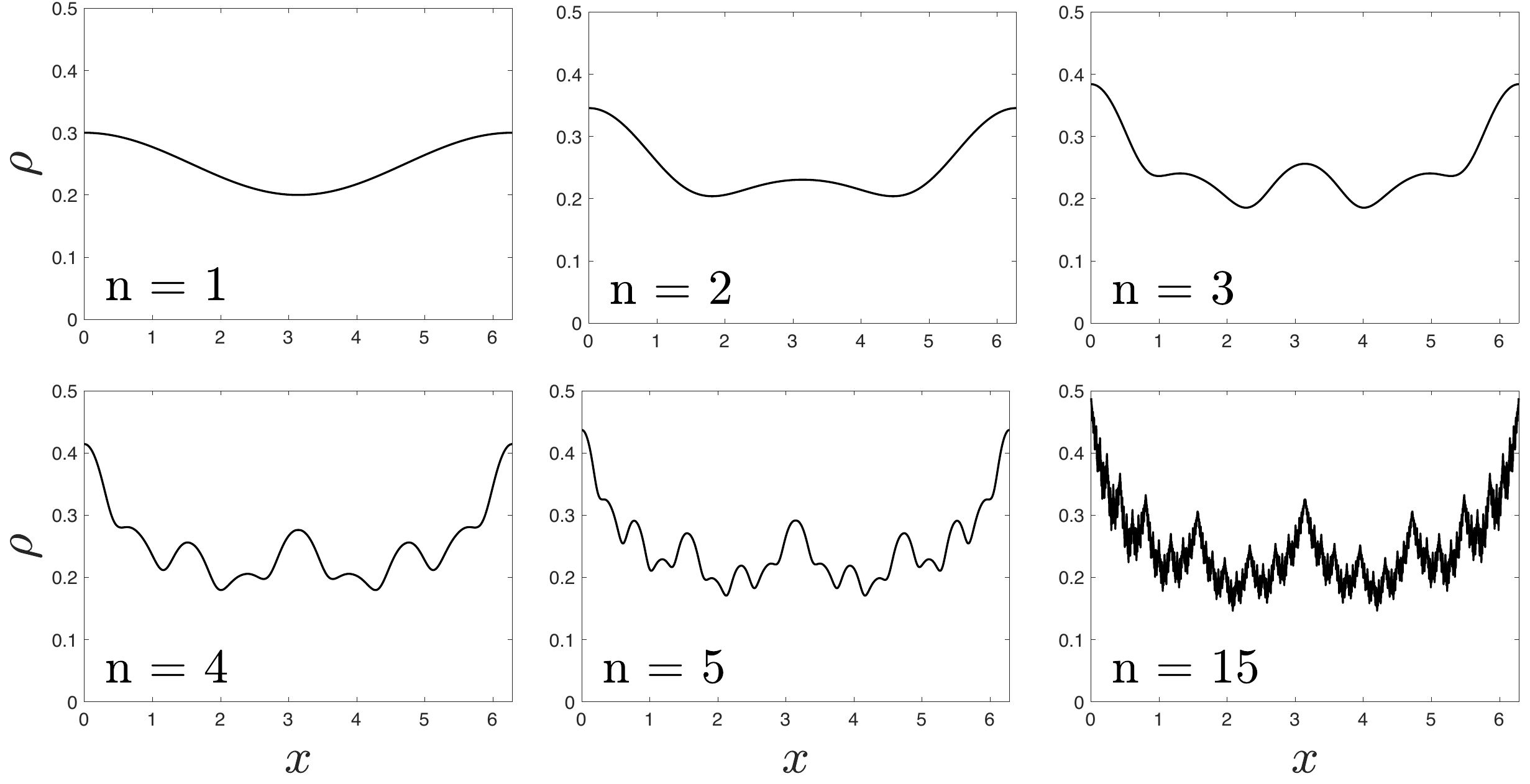} 
\caption{\label{fig2} The density $\rho$ after each iterate $n$ of the
  map $M$ with $r= 0.25$, $L_0 = 2 \pi$, and density variation given
  by Eq.~(\ref{r18}) with $B = 0.2$. More and more fine-scale
  structure develops at each iterate.  By $n = 15$ the density has
  visually converged to an invariant state.}
\end{figure}

\begin{figure}
\includegraphics[width = 0.7\columnwidth]{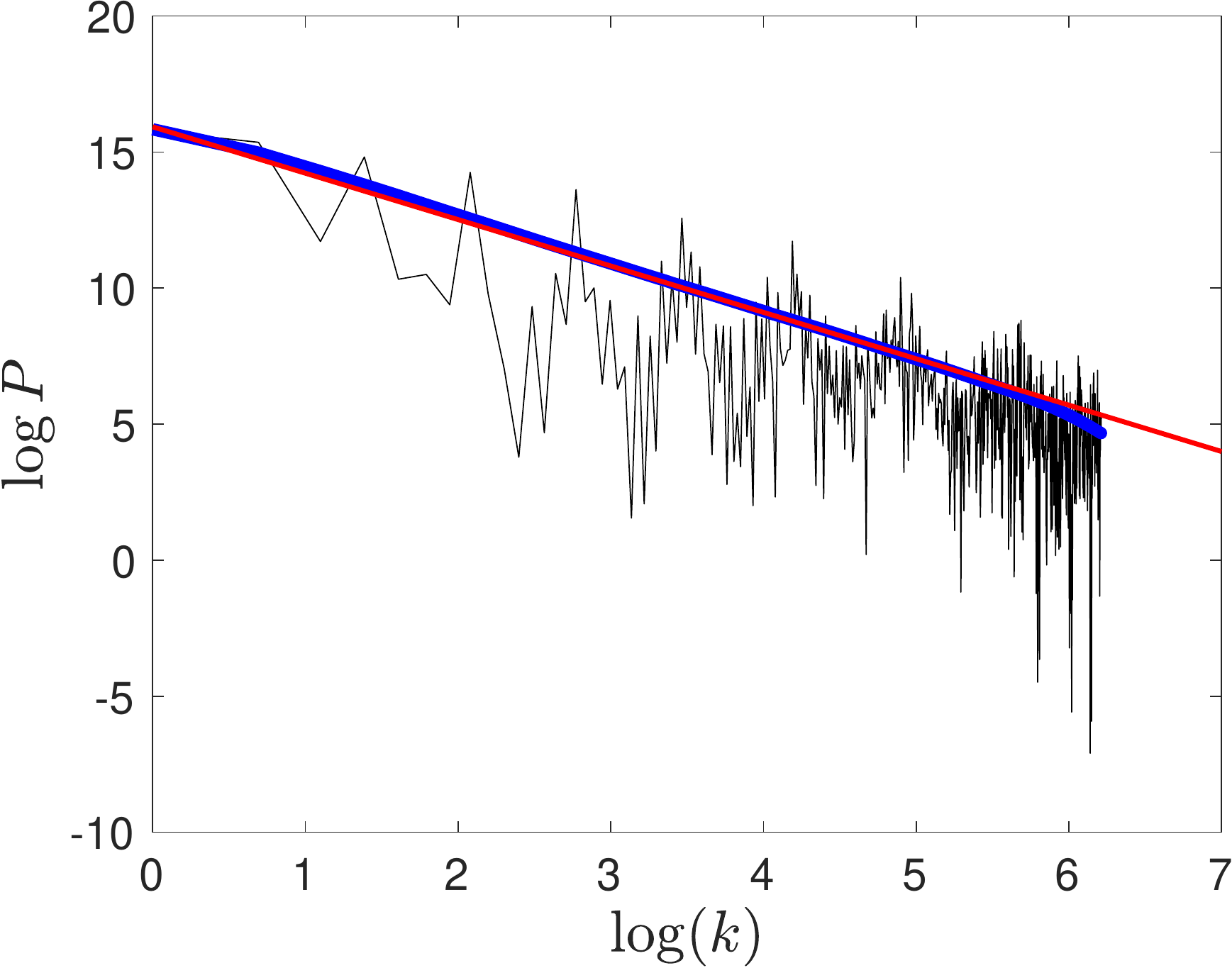}
\caption{\label{fig3} The power spectrum (black) of the final graph in
  Fig.~\ref{fig2} ($n = 15$).  The blue curve is the power spectrum
  smoothed according to Eq.~(\ref{r19}).  The smoothed spectrum is fit
  to a line (red) with slope -1.71.}
\end{figure}

\begin{figure}
\includegraphics[width = 1\columnwidth]{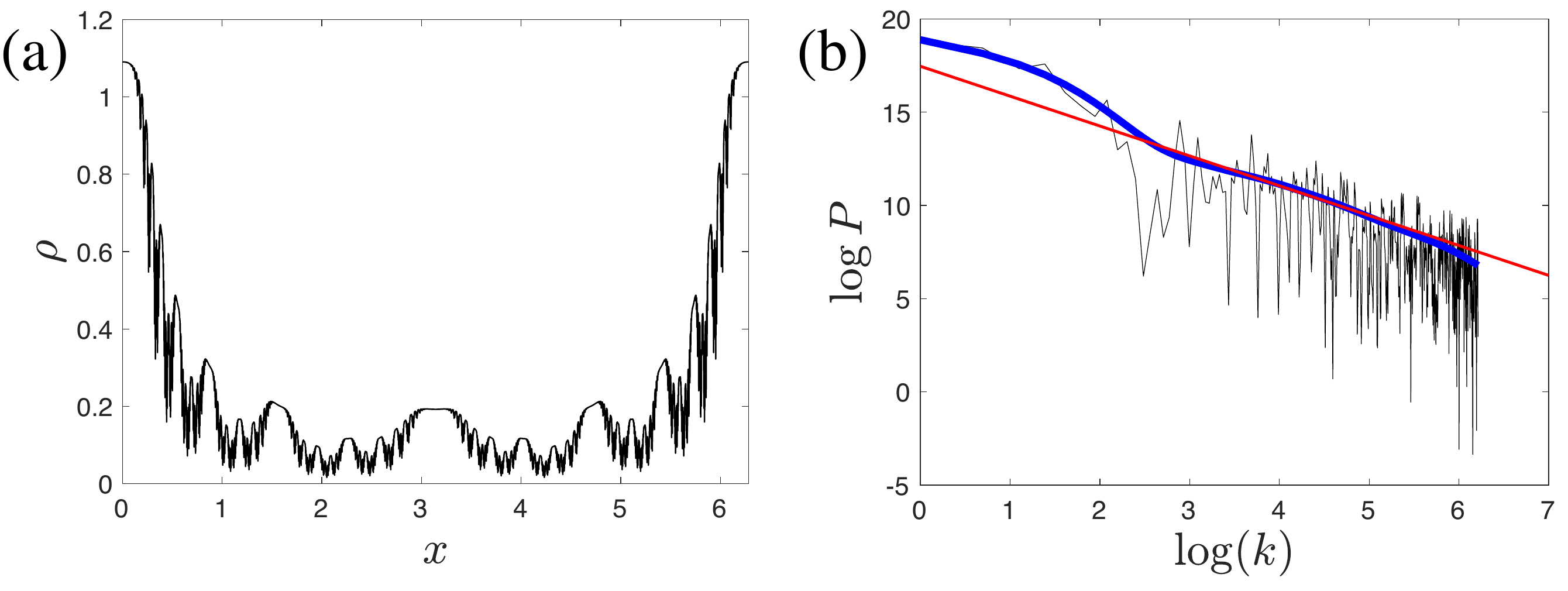}
\caption{\label{fig:LargeAmp} a) The stationary density for the same
  parameters as Fig.~\ref{fig2}, except $B = 0.7$.  b) The
  corresponding power spectrum.  The fit slope $-1.60$ differs
  slightly from Fig.~\ref{fig3}.}
\end{figure}

The sequence of compression by a factor of two, folding, and
large-scale density variations generates a mapping $M$ of an original
density $\rho_0(x)$ to a density $\rho(x)$; though the folding step
can be ignored if $\rho(x)$ is even and $L_0$-periodic as discussed
above.  This mapping depends on the form $R(x)$ of the density
variations imposed and the value of $r$ in the compression step.  The
mapping $M$ can then be repeatedly applied to the density.  We
numerically observe that it converges to a stationary fractal
function, as seen in Fig.~\ref{fig2} for the density variation
\begin{equation}
R(x) = 1 + B \cos(x),
\label{r18}
\end{equation}
with $B = 0.2$.  Figure~\ref{fig3} shows the power spectrum (black) of the
fractal image.  The blue line is the power spectrum averaged over
log(k), and the red line is a linear fit to this average.
For comparison, Fig.~\ref{fig:LargeAmp} shows the stationary density
distribution and power spectrum for larger amplitude density
variations with $B = 0.7$.

\section{Linear analysis of fractal generation}

\label{sec:linear}

\subsection{Linearization of compression dynamics}

The constant density 
\begin{equation}
\rho_0(x,t) = r
\end{equation}
and time-invariant velocity 
\begin{equation}
v_0(x,t) = (r-1)x
\label{r8}
\end{equation}
are steady state solutions of Eqs.~(\ref{r5}) and (\ref{r6}).  Note
that we have chosen a frame in which $v = 0$ at the origin $x=0$.
Expanding about this solution using 
\begin{align}
\rho & = \rho_0 + \epsilon g, \label{r14} \\
v & = v_0 + \epsilon f,
\end{align}
where $\langle g \rangle = 0$, we find to first order in $\epsilon$
\begin{align}
 \frac{\partial}{\partial t} g &=  (1-r)x  
  \frac{\partial}{\partial x} g  -r g,  \label{r9} \\
\frac{\partial f}{\partial x} & = g.  
\end{align}
Note that the density perturbation $g$ decouples from $f$.  Direct
substitution shows that the solution to Eq.~(\ref{r9}) for a given
initial condition $g(x,0)$ has the form
\begin{equation}
g(x,t) = h(t)g(x/\ell(t), 0),
\label{r23}
\end{equation}
where
\begin{align}
\ell(t) &= \exp( (r-1) t), \\
h(t) &= \exp(-rt). 
\end{align}
This implies that the Fourier transform $\tilde{g}(k,t)$ evolves in
time as
\begin{equation}
\tilde{g}(k,t) = h(t) \tilde{g}(k \ell(t), 0 ). 
\end{equation}
Thus as time evolves, the graph of the power spectrum in log-log space
shifts to larger wavenumbers and smaller amplitudes by translating
along a line of slope $-2r/(r-1)$.  After time $T$ given by
Eq.~(\ref{r11}), we find
\begin{equation}
\tilde{g}(k,T) = 2^{- r/(1-r) } \tilde{g}(k/2, 0). 
\label{r13}
\end{equation}
This equation is valid when the Fourier transform is smooth, but must
be treated carefully when the Fourier transform is singular, as when
$\rho$ is $L_0$-periodic at $t = 0$ and $t = T$.  In this case
$\tilde{g}(k,t)$ is zero except when $k = 2\pi n/L_0$ for
$n = 0, 1, 2, 3, ...$, at which value it is a delta function whose
amplitude we denote $\tilde{g}_n(t)$.  These coefficients satisfy
\begin{equation}
\tilde{g}_n(T) = 
\left\{ 
\begin{array}{ll}
2^{- r / (1-r) }  \tilde{g}_{ n/2 }(0), & n \text{
  even} \\
0, &  n \text{ odd.} 
\end{array}
\right.
\label{r15}
\end{equation}
Thus, the spacing between nonzero values of $\tilde{g}_n$ has
increased by a factor of 2 over time $T$.  Thus, when smoothed over
$n$, as discussed in the Appendix, $|\tilde{g}_n|^2$ scales as
\begin{align}
\langle |\tilde{g}_n(T)|^2 \rangle 
& \approx 0.5 \times  
2^{- 2r/(1-r)  }
\langle |\tilde{g}_{\lfloor n/2 \rfloor } (0)|^2 \rangle \nonumber \\
& \approx 
2^{- (1 + r)/(1-r)  }
\langle |\tilde{g}_{\lfloor n/2 \rfloor } (0)|^2 \rangle,
\label{r17}
\end{align}
where the angled brackets denote the average over $n$ and
$\lfloor \quad \rfloor$ denotes the floor function.  From this we see
that the smoothed power spectrum $\langle |\tilde{g}_n(t)|^2 \rangle$
slides down a line of slope $(1+r)/(1-r)$ in log-log space.

\subsection{Linearization of new density variations}

We assume the density variations introduced by Eq.~(\ref{r12}) scale 
as $\epsilon$, i.e.
\begin{equation}
R(x) = 1 + \epsilon p(x),
\label{r30}
\end{equation}
with $\langle p(x) \rangle = 0$.  Then, to first order in $\epsilon$
and using the expansion Eq.~(\ref{r14}), Eq.~(\ref{r12}) becomes
\begin{equation}
g(x) \mapsto g(x) + p(x),
\end{equation}
which in terms of the Fourier coefficients, we write as
\begin{equation}
\tilde{g}_n \mapsto \tilde{g}_n + \tilde{p}_n.
\end{equation}

\subsection{Linearized fractal generation}
\label{sec:linearfractal}

\begin{figure}
\includegraphics[width = .7\columnwidth]{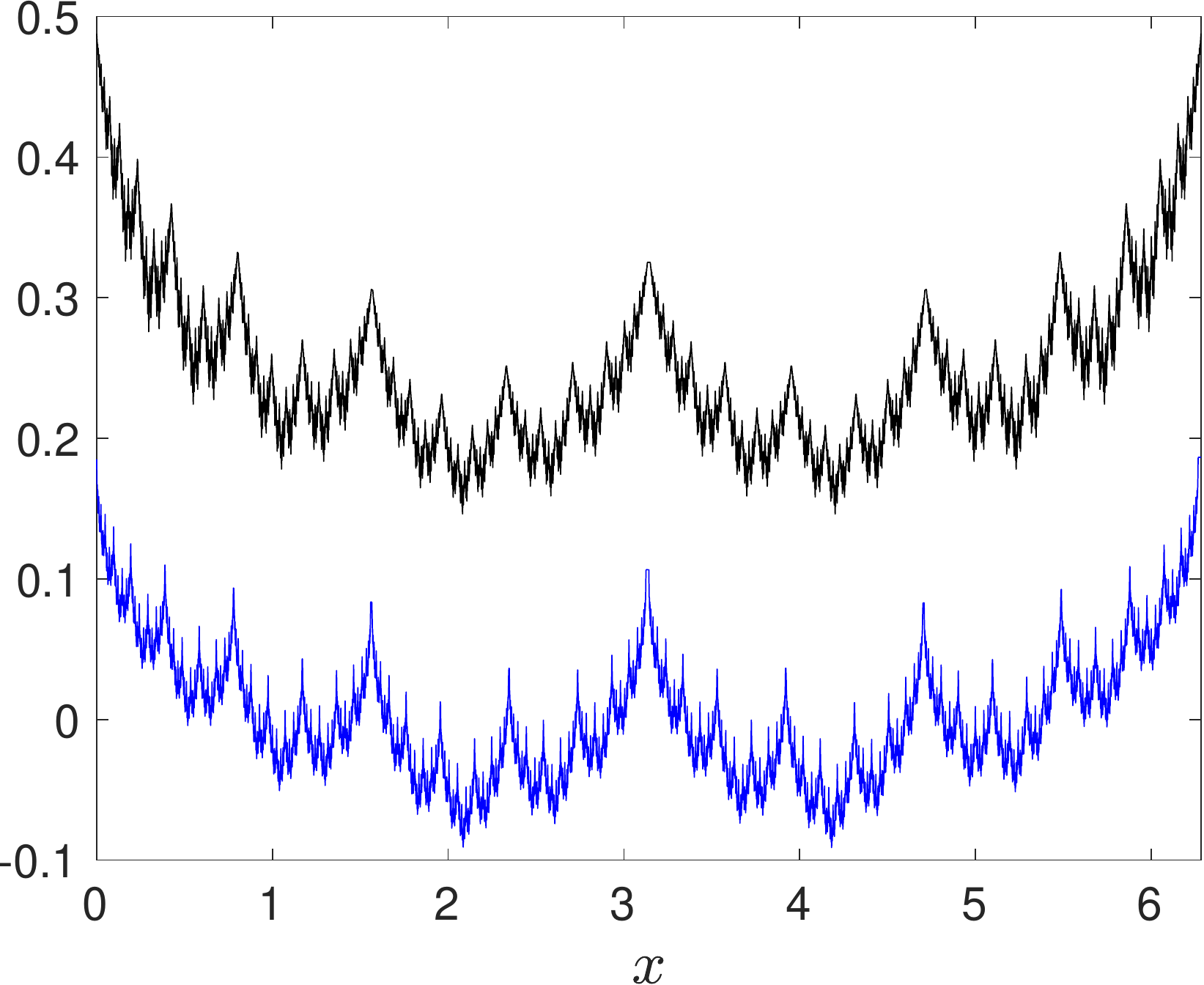} 
\caption{\label{fig4} The invariant function (blue) obtained from
  repeated application of the linearized dynamics Eq.~(\ref{r24}),
  with $r = 0.25$ and $L_0 = 2\pi$.  For comparison is the invariant
  function (black) obtained from the nonlinear dynamics and shown
  previously in Fig.~\ref{fig2}.}
\end{figure}

We denote by $A$ the linear operator that evolves $g(x,0)$ forward for
time $T$ according to Eq.~(\ref{r23}).  Then the total linearized
dynamics is
\begin{equation}
g \mapsto A g+ p.
\label{r24}
\end{equation}
Repeated application of this map, beginning with $g = 0$ and using
\begin{equation}
p(x) = 0.2 \cos(x),
\label{r25}
\end{equation}
converges to the stationary function (blue) in Fig.~\ref{fig4}, which
is consistent with the nonlinear stationary density from
Fig.~\ref{fig2} and reproduced in Fig.~\ref{fig4} (black).  It is
straightforward to see that
\begin{equation}
P = \sum_{m = 0}^\infty A^m p, 
\end{equation}
is the unique invariant function of Eq.~(\ref{r24}) and that  any
initial function will converge to it.

It was noted to us by S.~Berman that when $p(x)$ is a cosine, as in
Eq.~(\ref{r25}), the invariant function $P(x)$ is the Weierstrass
function, which was the first published example of a continuous
function that is nowhere differentiable.  Furthermore, fixed-point
equations with  the form of Eq.~(\ref{r24}) were introduced by de Rham to
characterize such singular functions.  Such techniques have a history
of applications to dynamical systems, e.g. Ref.~\onlinecite{Tasaki98}.

In Fourier space, the linear operator $\tilde{A}$ that maps
coefficients $\tilde{g}_n(0)$ forward to $\tilde{g}_n(T)$ is given by
Eq.~(\ref{r15}), and the invariant function is
\begin{equation}
\tilde{P}  = \sum_{m = 0}^\infty \tilde{A}^m \tilde{p}, 
\end{equation}
with coefficients
\begin{equation}
\tilde{P}_n = \sum_{m = 0}^\infty  2^{- m r/(1-r)}
              \tilde{p}_{n/2^m}, 
\label{r16}
\end{equation}
where it is understood that $\tilde{p}_{n/2^m} = 0$ if $n/2^m$ is not
an integer.

\begin{figure}
\includegraphics[width = .7\columnwidth]{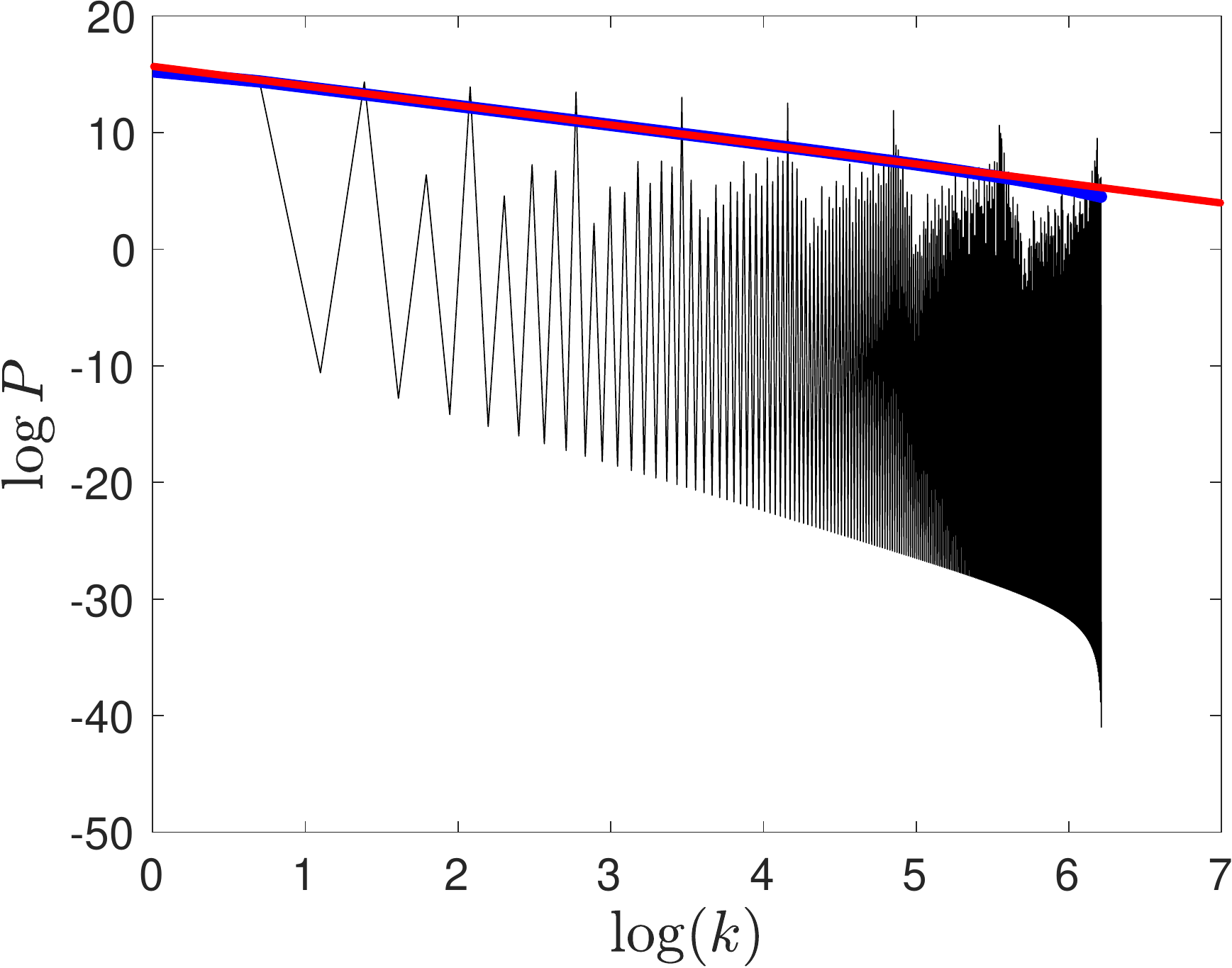} 
\caption{\label{fig5} The power spectrum (black) of the blue curve in
  Fig.~\ref{fig4}.  The blue curve above is the power spectrum
  smoothed according to Eq.~(\ref{r19}).  The smoothed spectrum is fit
  to a line (red) with slope -1.67.}
\end{figure}

Assuming that the density variation function $p(x)$ is analytic, its
Fourier coefficients $\tilde{p}_n$ decay exponentially in $n$, meaning
that the sum in Eq.~(\ref{r16}) is dominated by the first nonzero
term.  This term will have an $m$ value satisfying  $m \approx \log(n)/\log(2)$,
and hence
\begin{equation}
\tilde{P}_n \propto n^{-r/(1-r)}, 
\end{equation}
from which follows
\begin{equation}
|\tilde{P}_n|^2 \propto n^{-2r/(1-r)}, 
\end{equation}
and finally
\begin{equation}
\langle |\tilde{P}_n|^2 \rangle \propto \frac{1}{n} n^{-2r/(1-r)} = n^{-(1+r)/(1-r)}, 
\end{equation}
where the factor of $1/n$ that appears upon averaging is due to the
diminishing density of nonzero coefficients, as discussed prior to
Eq.~(\ref{r17}).  Thus in the linearized model, we  find 
\begin{equation}
\beta = \frac{1+r}{1-r}.
\label{r20}
\end{equation}
This is the same scaling that we previously identified for the
temporal behavior in Eq.~(\ref{r17}), highlighting the origin of this
scaling in the compression dynamics.

\begin{figure}
\includegraphics[width = .7\columnwidth]{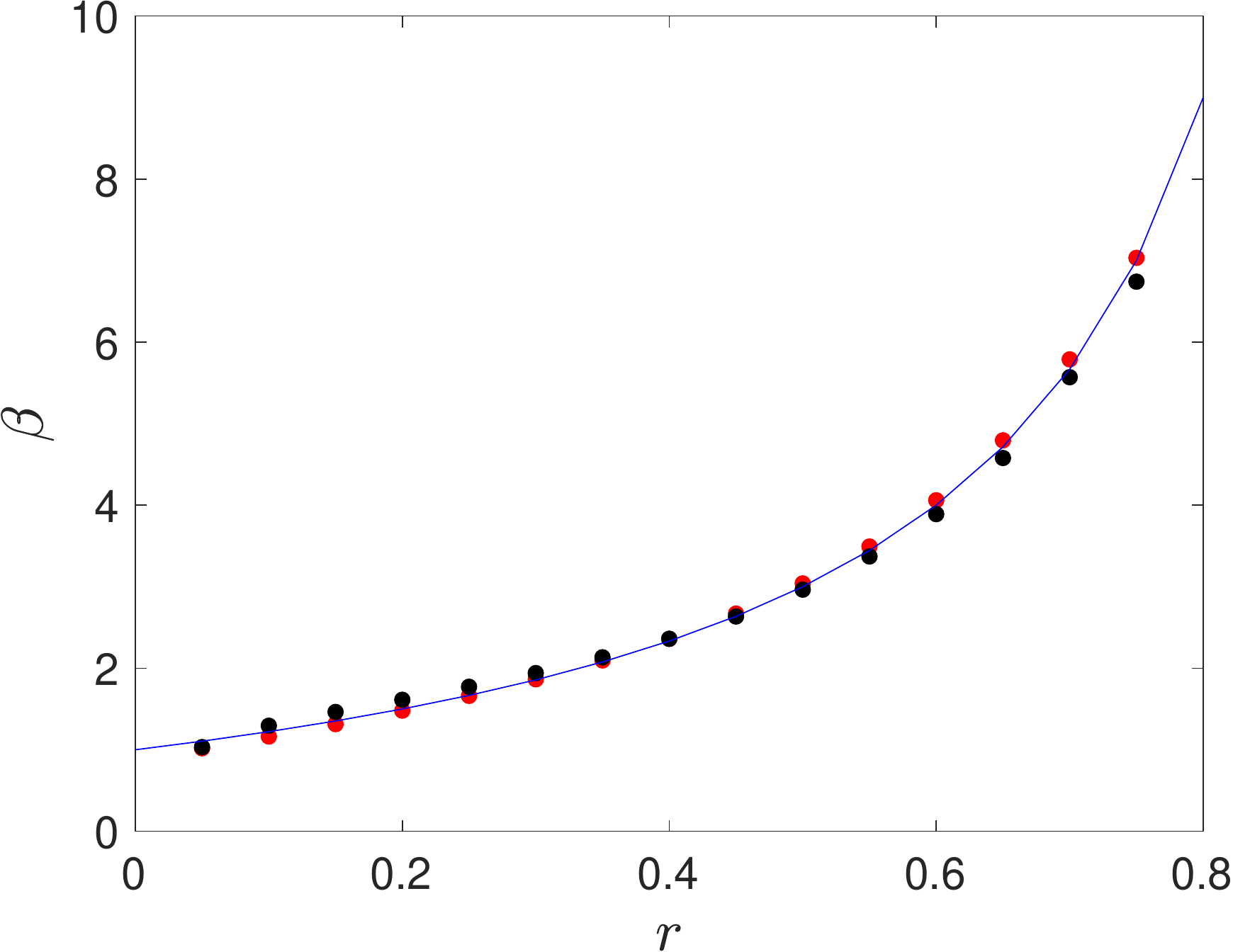} 
\caption{\label{fig6} The variation of $\beta$ with $r$.  The black
  dots are computed numerically from the nonlinear analysis with
  $B = 0.2$.  The red dots are computed numerically from the linear
  analysis.  The blue curve is the graph of Eq.~(\ref{r20}).}
\end{figure}

To check this formula numerically, we plot in Fig.~\ref{fig5} the
power spectrum of the blue curve in Fig.~\ref{fig4}, which has
$r = 0.25$.  The linear fit yields a slope of $-1.67$, which agrees
with the value $\beta = 1.67$ obtained from Eq.~(\ref{r20}).
Figure~\ref{fig6} provides a more comprehensive comparison of the
nonlinear analysis, the linear analysis, and Eq.~(\ref{r20}).  The black
and red dots show $\beta$ versus $r$ for the nonlinear and linearized
systems, respectively, using $R$ and $p$ given by Eqs.~(\ref{r18})
(with $B = 0.2$) and (\ref{r25}).  The data are quite consistent with
each other.  The blue curve is the graph of Eq.~(\ref{r20}), and it
nicely tracks the red data points.

%(* It is also independent of the timing of the folds.  For example, you
%could fold after factor of 3, etc.   *)

\subsection{The fluctuation injection scale}

\begin{figure}
\includegraphics[width = .7\columnwidth]{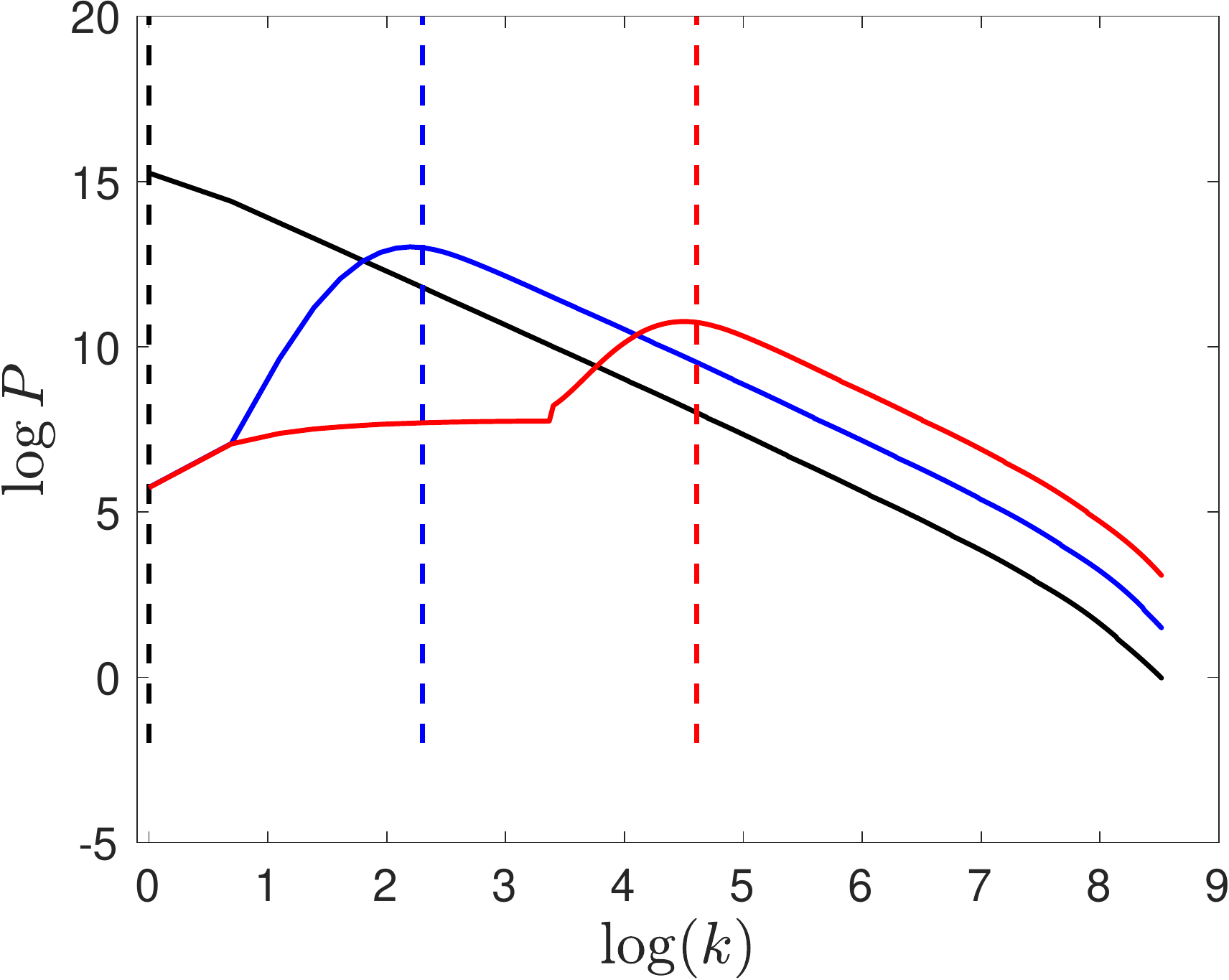} 
\caption{\label{fig:injection} The (smoothed) power spectra of the
  linearized dynamics for three different sinusoidal forms for the
  injected density fluctuations.  The black curve uses an injection
  wavelength equal to the full width of the material.  The red and
  blue curves use wavelengths that are a factor of 10 and 100 smaller,
  respectively.  The dashed vertical lines denote the scale of the
  injected density fluctuations.  Note that the linear behavior only
  occurs when $k$ is greater than the injection value.}
\end{figure}

The value of $\beta$ does not depend on the details of the density
fluctuations introduced by $p(x)$, except that its Fourier
coefficients should fall off sufficiently rapidly.  However, the
$k^{-\beta}$ scaling only begins at length scales below the smallest
length scale of $p(x)$.  Figure~\ref{fig:injection} illustrates this
by plotting the smoothed power spectrum for $p(x)$ with wavelengths
equal to the width of the block of material divided by 1
[$p(x) = \cos(x)$, black], 10 [$p(x) = \cos(10 x)$, blue], and 100
[$p(x) = \cos(100 x)$, red].  These wavelengths are the scales at which
density fluctuations are injected into the system.  The $k$ value for
each wavelength is denoted by the dashed vertical line of the
corresponding color.  The linear behavior only begins to the right of
each line, i.e. at length scales smaller than the injection scale.

\section{The special case of incompressibility in Step 1}

\label{sec:incompressible}

Returning to Eq.~(\ref{r4}) and taking the limit $\rho_m = \infty$ and
setting $\alpha = \gamma$, density is conserved in the Lagrangian
frame,
\begin{equation}
\frac{D}{Dt} \rho(x,t) = 0,
\end{equation}
so that the material is incompressible, regardless of its density.
Furthermore, Eq.~(\ref{r2}) reduces to 
\begin{equation}
\frac{\partial v}{\partial x} = -\alpha,
\end{equation}
so that the system is uniformly compressed in the $x$ direction.
Combining this with the density variations in Step 3, Eq.~(\ref{r12}),
we obtain a steady state fractal density.  (We ignore subtle points
about the convergence of this function, which is technically a
distribution.)  This dynamics is admittedly physically inconsistent, as
we cannot imagine a situation in which density fluctuations could
arrise via folding on the active scale (Step 3) without the material
also being subject to compression or expansion in Step 1.
Nevertheless, it is instructive to compute the resulting value of
$\beta$, which is straightforward when linearizing the density
fluctuations, i.e. using Eq.~(\ref{r30}).  Repeating the analysis in
Sec.~\ref{sec:linearfractal}, one finds
\begin{equation}
\tilde{P}_n = \sum_{m = 0}^\infty  \tilde{p}_{n/2^m}, 
\end{equation}
which simply reflects the fact that as a cosine density fluctuation is
squeezed in $x$, its amplitude remains constant.  This is the same
result as Eq.~(\ref{r16}) with $r = 0$ and therefore gives the same
power spectrum decay with $\beta = 1$.  Thus, even if one were to
assume incompressibility in Step 1, the resulting value of $\beta$
disagrees with the experimental measurements in Fig.~\ref{fig9} (at
least within the linearized analysis.)

\section{Conclusions}

\label{sec:conclusions}

By using a simple model of fractal generation, we have highlighted the
necessity both of introducing density fluctuations on large scales and
of a variable compressibility in the material.  At least within the
linearized analysis, the details of the density fluctuations are
irrelevant to the value of $\beta$.  The density fluctuations can take
any functional form, so long as they have a minimum length scale.  The
$k^{-\beta}$ scaling then manifests at scales below this smallest
length scale.  In the case of the microtubule-based active nematic
material, the injection scale of the density fluctuations is the
active length scale of the system, i.e. the average defect spacing.
Physically, the injected density variations are most naturally
explained by the creation of defect pairs and the associated
fracturing.

The $\beta$ parameter depends on the compressibility via $r$, and in
the linearized analysis we derived an explicit relationship,
Eq.~(\ref{r20}).  It is tempting to thus translate the experimentally
measured values of $\beta$ (Fig.~\ref{fig9}) into a material parameter
$r$ via
\begin{equation}
r = \frac{\beta - 1}{\beta + 1},
\end{equation}
yielding values of $r$ in the range $0.1$ -- $0.3$.  But what is the
physical meaning of $r$?  Within the context of the simple model,
we can  combine Eqs.~(\ref{r27}) and (\ref{r26}), to rewrite $r$ as
\begin{equation}
r = 1 - \gamma/\alpha.
\end{equation}
Recall $\gamma$ is the (uniform and constant) extension rate and
$\alpha$ can be interpreted as the transverse compression rate in the
low-density limit [Eq.~(\ref{r2})].  Whether this interpretation from
the simple fractal model will hold for the actual laboratory system
is not clear.  This issue could be addressed via a realistic 2D
hydrodynamic model with a density-dependent compressibility.
Specifically, the dependence of $\beta$ on the material properties and
system parameters would make an interesting and informative study.

Finally, it would be interesting to identify whether other physical
systems exhibit a similar mechanism for fractal formation.  Such
systems would exhibit fractal structure not in the patterning of
impurities mixed into the system, but in morphological properties such
as density or surface texture.  The classic example of chaos in a
taffy puller comes to mind~\cite{Thiffeault18}, in which large-scale surface
irregularities are introduced periodically by the folding of the taffy
and then pushed down to smaller length scales by the stretching and
compression dynamics.

\begin{acknowledgments}
The authors acknowledge generous funding from the National Science
Foundation, through several awards: DMR-1808926, NSF-CREST: Center for
Cellular and Biomolecular Machines at UC Merced (HRD-1547848), and
from the Brandeis Biomaterials facility MRSEC-1420382, which provided
materials.  We would also like to thank Simon Berman for critiquing the
manuscript, and in particular for pointing out the work in
Ref.~\onlinecite{Tasaki98}.
\end{acknowledgments}

\section*{DATA AVAILABILITY}

The data that support the findings of this study are available from the corresponding author upon reasonable request.

\appendix

\section{Spectral averaging}

We smooth out local variations in a function $f(k)$ using a Gaussian
function of width $\sigma$ in log space
\begin{equation}
\langle f \rangle(k) = \int_0^\infty f(k') \frac{1}{N(k)}\exp \left[ - \frac{1}{2 \sigma^2}
  (\log k' - \log k)^2 \right] dk',
\label{r19}
\end{equation}
with  normalization 
\begin{align}
N(k) & = \int_0^\infty  \exp \left[ - \frac{1}{2 \sigma^2}
  (\log k' - \log k)^2 \right] dk' \nonumber \\
& = \sqrt{2 \pi} \sigma e^{\sigma^2/2} k.
\end{align}
Note that this smoothing preserves the $L^1$-norm, i.e.
$\int_0^\infty |\langle f \rangle (k)| dk = \int_0^\infty | f(k) |
dk$.  Thus, spectral averaging of the power function preserves the
total power.  For all computations in this paper, we use $\sigma = 0.4$.  
% A straightforward computation also shows that if
% $f(k) = g(k/c)$, then
% $\langle f \rangle (k) = \langle g \rangle (k/c)$, for any $c>0$.
% This shows that the spectral averaging is scale independent.

%\bibliographystyle{aipnum4-1}
%\bibliography{/Users/kevinm/docs/misc/MyBibDeskBib}


\end{document}